%% file: rev_paper.tex
\documentclass[12pt, twoside]{article}

\usepackage{a4wide, amsmath, latexsym, epsfig,
  psfrag,graphicx, rotating, fancyhdr, tabularx, afterpage}

\newcommand{\id}{{\rm 1\kern-.12em
\rule{0.3pt}{1.5ex}\raisebox{0.0ex}{\rule{0.1em}{0.3pt}}}}

\begin{document}
\include{titlepage}

\section{Introduction}


The next generation of high energy colliders will permit a discovery 
of supersymmetric particles and accurate measurements of their 
properties \nolinebreak \cite{futcoll}. From precise determination of 
masses, cross sections and asymmetries 
the fundamental para\-meters of the underlying theory \nolinebreak
\cite{susysuche} can be reconstructed. This knowledge  
will provide insight into the supersymmetry breaking mechanism and its
relation to Grand Unification. 

The reconstruction of the basic SUSY-para\-meters from the experimental
data requires reliable relations between physical observables and
fundamental para\-meters. Higher-order corrections to tree-level relations
have to be calculated and to be taken into account when transforming
from physical para\-meters to the fundamental ones. 

In this paper, we derive the one-loop contributions to the sfermion-mass
spectrum. An on-shell renormalization scheme is applied where all
masses are treated as pole masses. Due to $SU(2)$-symmetry, the
soft-breaking para\-meters of the supersymmetric partners  of the
left-handed fermions are identical. Hence, in each generation of up- and
down-type sfermions one
sfermion-mass is dependent on the remaining masses in that
generation. Therefore, on one-loop level the pole mass of that sfermion
receives a shift with respect to its tree-level value. This shift has been
calculated including the complete set of one-loop
diagrams. As a by-product, counter\-terms for the soft-breaking para\-meters of
the sfermionic sector
are derived and are being implemented into the MSSM version of {\ttfamily
  FeynArts} \nolinebreak \cite{HaScha}. Since
these counterterms are specified for the basic breaking parameters,
they are different from those of~\cite{GuSoHo},
where another way of renormalization has been performed that
consists of introducing counterterms directly for the physical parameters, 
i.e.\ masses and mixing angles
instead of the soft-breaking parameters.

Beginning with a review of the sfermion-mass matrix on Born level in
section \nolinebreak \ref{bornlevel}, renormalization conditions are
specified and explicit mass counter\-terms are calculated in 
section~\ref{rencond}. In section \ref{numres}, we present our numerical
results.  

\section{The scalar fermion sector at the Born level}\label{bornlevel}

In the MSSM, supersymmetry breaking is implemented by explicitly adding 
soft-breaking terms to the symmetric Lagrangian. The sfermion-mass
terms of the Lagrangian, for a given species of sfermions $\tilde{f}$,
 can be written as the bilinear expression 
\begin{align}
{\mathcal{L}}_{\tilde{f}\text{-mass}} &= - \begin{pmatrix}
{{\tilde{f}}_{L}}^{+}, {{\tilde{f}}_{R}}^{+} \end{pmatrix}
\mathcal{M}_{\tilde{f}}\begin{pmatrix}{\tilde{f}}_{L}\\{\tilde{f}}_{R}
\end{pmatrix} 
\end{align} 
with $\mathcal{M}_{\tilde{f}}$ as the sfermion-mass matrix squared,  
\begin{align}\label{Sfermionmassenmatrix}
\mathcal{M}_{\tilde{f}} = \begin{pmatrix}  m_f^2 + M_L^2 + M_Z^2 c_{2
  \beta} (T_f^3 - Q_{{f}} s_W^2) & m_f (A_f^{*} - \mu \kappa) \\
  m_f (A_f - {\mu}^{*} \kappa) &   m_f^2 + M_{{\tilde{f}}_R}^2 + M_Z^2
  c_{2 \beta} Q_{{f}} 
  s_W^2 \end{pmatrix},
\end{align}
where the quantities $M_L^2$, $M_{{\tilde{f}}_R}^2$ and $A_f$ 
denote the soft-breaking
para\-meters. In this paper we treat these parameter as real
quantities. $\tan \beta = \frac{v_2}{v_1}$ denotes the ratio of the 
Higgs vacuum expectation values $v_1$ and $v_2$, and $\mu$ is the 
supersymmetric Higgs mass parameter.
As abbreviations, $c_{2 \beta} = \cos (2 \beta)$ and $s_W = \sin
{\theta}_W$ are used where ${\theta}_W$ is the weak mixing
angle. The para\-meter $\kappa$ is defined as
$\kappa = \cot \beta$ for up-type
squarks and $\kappa = \tan \beta$ for down-type squarks and
electron-type sleptons. $m_f$ is the mass of the fermion $f$,
$Q_{{f}}$ the 
electromagnetic charge, and  $T_f^3$ the
isospin of $f$.

As far as we do not consider right-handed neutrinos within the MSSM, 
corresponding superpartners do not exist. 
Thus, for sneutrinos the sfermion-mass matrix
is 1-dimensional, with only the left-handed entry of \eqref{Sfermionmassenmatrix}.

The mass matrix \eqref{Sfermionmassenmatrix}
can be diagonalized by a transformation of the
$\tilde{f}_{L,R}$ fields with the help of a unitary matrix
${\mathcal{U}}_{\tilde{f}}$,
\begin{align}\label{transformation}
\begin{pmatrix}{\tilde{f}}_{1}\\{\tilde{f}}_{2} \end{pmatrix} =
  {\mathcal{U}}_{\tilde{f}} \begin{pmatrix}{\tilde{f}}_{L}\\{\tilde{f}}_{R}
\end{pmatrix} \qquad \begin{pmatrix}{\tilde{f}}_{L}\\{\tilde{f}}_{R}
\end{pmatrix} = 
  {\mathcal{U}}_{\tilde{f}}^{+}
  \begin{pmatrix}{\tilde{f}}_{1}\\{\tilde{f}}_{2} \end{pmatrix}\,. 
\end{align}
In our case of real parameters, ${\mathcal{U}}_{\tilde{f}}$  
can be parameterized in terms of a mixing angle 
$\theta_{\tilde{f}}$,
\begin{align}\label{transfmatrix}
{\mathcal{U}}_{\tilde{f}} = \begin{pmatrix} \cos \theta_{\tilde{f}} &
  \sin \theta_{\tilde{f}} \\ - \sin \theta_{\tilde{f}} & \cos
  \theta_{\tilde{f}} \end{pmatrix}\,.
\end{align}
In the $(1, 2)$-basis, the squared-mass matrix is diagonal,
\begin{align}
{\mathcal{D}}_{\tilde{f}} &= {\mathcal{U}}_{\tilde{f}}
{\mathcal{M}}_{\tilde{f}}{\mathcal{U}}_{\tilde{f}}^{+} = \begin{pmatrix}
m_{{\tilde{f}}_1}^2 & 0 \\ 0 & m_{{\tilde{f}}_2}^2 \end{pmatrix}\,,
\end{align}
with the eigenvalues $m_{{\tilde{f}}_1}^2$ and $m_{{\tilde{f}}_2}^2$ given by
\begin{align}
\begin{split}\label{Sfermionmassenmatrixeigenwerte}
&m_{{\tilde{f}}_{1,2}}^2 = \frac{1}{2}(M_L^2
+M_{{\tilde{f}}_R}^2) +  m_f^2 + \frac{1}{2}T_f^3 M_Z^2 c_{2 \beta} \\&
\qquad\ \quad \pm
\frac{1}{2} \sqrt{\bigl[M_L^2 - M_{{\tilde{f}}_R}^2 + M_Z^2 c_{2
  \beta} (T_f^3 - 2 Q_{{f}} s_W^2)\bigr]^2 + 4  m_f^2 |A_f - {\mu} \kappa|^2}.
\end{split}
\end{align}

\section{The scalar fermion sector at the one-loop level}\label{rencond}

For renormalization of the sfermion sector,
counterterms for the mass matrix \eqref{Sfermionmassenmatrix} are introduced,
\begin{align}\label{Sfermionparameterrenormierung}
 \mathcal{M}_{\tilde{f}} &\rightarrow
  \mathcal{M}_{\tilde{f}} + \delta \! \mathcal{M}_{\tilde{f}},
\end{align}
where $\delta \! \mathcal{M}_{\tilde{f}}$ contains the counterterms of the parameters
appearing in \eqref{Sfermionmassenmatrix}.

By field renormalization, the sfermion fields are replaced by
renormalized fields and $Z$-factors, 
\begin{align}\label{Sfermionfeldrenormierung}
\begin{pmatrix}{\tilde{f}}_{L}\\{\tilde{f}}_{R} \end{pmatrix}
  &\rightarrow (\id + \frac{1}{2} \delta \!
  {\mathcal{Z}}_{\tilde{f}})
  \begin{pmatrix}{\tilde{f}}_{L}\\{\tilde{f}}_{R} \end{pmatrix}
  \quad \text{with} \quad
  \delta \! {\mathcal{Z}}_{\tilde{f}} = \begin{pmatrix}\delta \!  {Z_{\tilde{f}}}_L & 0 \\ 0 &
  \delta \! {Z_{\tilde{f}}}_R
  \end{pmatrix}.
\end{align}
This assignment
forms the minimal set of renormalization constants 
satisfying the symmetry relations \nolinebreak \cite{HoKrausStoe}, 
and is sufficient to absorb all the divergencies.

The renormalization transformations 
\eqref{Sfermionparameterrenormierung} and \eqref{Sfermionfeldrenormierung}, 
together with \eqref{transformation}, yield the renormalized sfermion 
self-energies ${\hat{\Sigma}}_{\tilde{f}}$
from the unrenormalized ones, $\Sigma_{\tilde{f}}$, according to 
\begin{align}\label{renSelbstenergie}
{\hat{\Sigma}}_{\tilde{f}}(k^2) = \Sigma_{\tilde{f}} (k^2) + k^2 \delta
\! \tilde{\mathcal{Z}}_{\tilde{f}} - 
\frac{1}{2}(\delta \! \tilde{\mathcal{Z}}_{\tilde{f}} \mathcal{D}_{\tilde{f}} +
\mathcal{D}_{\tilde{f}} \delta \! \tilde{\mathcal{Z}}_{\tilde{f}})- \mathcal{U}_{\tilde{f}}
\delta \! \mathcal{M}_{\tilde{f}} \mathcal{U}^{+}_{\tilde{f}}.
\end{align}
Thereby, $\Sigma_{\tilde{f}}$ denotes the matrix of the diagonal and non-diagonal self-energies
for $\tilde{f}_{1,2}$.
$\delta \! \tilde{\mathcal{Z}}_{\tilde{f}}$ is used as an
abbreviation, $\delta \! \tilde{\mathcal{Z}}_{\tilde{f}} =
\mathcal{U}_{\tilde{f}} \delta \! \mathcal{Z}_{\tilde{f}}
\mathcal{U}^{+}_{\tilde{f}}$.

It is convenient to introduce,
instead of \eqref{transformation},
a more general transformation at the one-loop level, replacing
\begin{align}\label{vollstTransfmatrix}
\mathcal{U}_{\tilde{f}} \rightarrow \mathcal{R}_{\tilde{f}} = (\id +
\frac{1}{2} \delta \! \mathcal{Z}_{U_{\tilde{f}}})\mathcal{U}_{\tilde{f}}\;,
\end{align}
with an additional UV-finite matrix $\delta \!
\mathcal{Z}_{U_{\tilde{f}}}$. This procedure yields  a
non-diagonal $Z$-matrix for the sfermion fields with four independent
entries.  
In that case
the renormalized
self-energies are given by
 \begin{align}
 \begin{split}\label{vollstrenSelbstenergie}
 {\hat{\Sigma}}_{\tilde{f}}(k^2) &= \Sigma_{\tilde{f}} (k^2) +
 \frac{1}{2} k^2 (\delta \! \breve{\mathcal{Z}}^{+}_{\tilde{f}} +
 \delta \! \breve{\mathcal{Z}}_{\tilde{f}}) -
 \frac{1}{2}(\delta \! \breve{\mathcal{Z}}^{+}_{\tilde{f}}
 \mathcal{D}_{\tilde{f}} + \mathcal{D}_{\tilde{f}} \delta \!
 \breve{\mathcal{Z}}_{\tilde{f}})-  \mathcal{U}_{\tilde{f}}
 \delta \! \mathcal{M}_{\tilde{f}} \mathcal{U}^{+}_{\tilde{f}} 
 \end{split}
 \end{align}
 with
 \begin{align}
 \delta \! \breve{\mathcal{Z}}_{\tilde{f}}=\mathcal{U}_{\tilde{f}}
 \delta \! \mathcal{Z}_{\tilde{f}} 
 \mathcal{U}_{\tilde{f}}^{+}- \delta \! \mathcal{Z}_{U_{\tilde{f}}} =
 \begin{pmatrix}\delta \! 
  \breve{Z}_{\tilde{f}_{11}} & \delta \! \breve{Z}_{\tilde{f}_{12}}
  \\ \delta \! \breve{Z}_{\tilde{f}_{21}} & \delta \!
  \breve{Z}_{\tilde{f}_{22}} \end{pmatrix}.
 \end{align}
This procedure, in analogy to the one for renormalization of the
chargino and neutralino sector performed in \cite{fh}, 
will be the basis of the forthcoming discussion.

\subsection{Renormalization conditions}

All independent para\-meters in the sfermion-mass matrix 
$\mathcal{M}_{\tilde{f}}$ 
in \eqref{Sfermionmassenmatrix} are replaced by  renormalized para\-meters
and the corresponding counter\-terms, which form the
counter\-term matrix $ \delta \mathcal{M}_{\tilde{f}}$. Only the
counter\-terms of the soft-breaking para\-meters $M_L^2$,
$M^2_{{\tilde{f}}_R}$ and $A_f$ 
have to be determined within the sfermion sector, the others follow from the
gauge, gaugino, Higgs and fermion sectors.

For one generation of squarks, neglecting mixing between generations,
there exists one mass matrix for the $u$-type squarks and one for
the $d$-type squarks. Because of $SU(2)$-invariance, the para\-meter
$M_{L_{\tilde{q}}}$ is the same for the $u$- and the $d$-type
squarks. Therefore, in 
one generation of squarks, there are five para\-meters $M_{L_{\tilde{q}}}^2$,
$M^2_{{\tilde{u}}_R}$, $M^2_{{\tilde{d}}_R}$, $A_u$, $A_d$ with
counter\-terms to be determined within the sfermion sector. Hence, five
renormalization conditions are required. 

On-shell mass-renormalization conditions can be imposed on both mass
eigenstates of either the $u$- or $d$-type sfermions. Here we choose 
the isospin``+'' system, with the on-shell conditions 
expressed in terms of the diagonal entries of (\ref{vollstrenSelbstenergie}),
\begin{align}\label{onshellBed1}
\text{Re}\, {\hat{\Sigma}}_{\tilde{u}_{ii}}(m_{{\tilde{u}}_i}^2) &= 0
\quad\ \text{with} \quad\ i=1,2\,.
\end{align}
For the $\tilde{d}$ system, the on-shell condition is imposed 
for the $\tilde{d}_2$-squarks,
\begin{align}\label{onshellBed2}
\text{Re}{\hat{\Sigma}}_{\tilde{d}_{22}}(m_{{\tilde{d}}_2}^2) &= 0 \, ,
\end{align}
as long as ${\tilde{d}}_2 \neq \pm {\tilde{d}}_L$. According to
\eqref{Sfermionmassenmatrixeigenwerte}, we have chosen the heavier
squark to be  ${\tilde{d}}_1$ and the lighter one to be 
${\tilde{d}}_2$, hence mixing angles $\theta_{\tilde{f}} >
|\frac{\pi}{4}|$ can occur in the matrix
$\mathcal U_{\tilde{f}}$ in \eqref{transfmatrix}. In case of ${\tilde{d}}_2
= \pm {\tilde{d}}_L$, corresponding to a mixing angle $\theta_{\tilde d}
= \mp \frac{\pi}{2}$, the self energy ${\hat{\Sigma}}_{\tilde{d}_{22}}$
contains only the counter\-term $\delta \! M_{L_{\tilde{q}}}^2$ which is
already fixed 
by one of the conditions \eqref{onshellBed1}. The renormalization
condition \eqref{onshellBed2} has to be replaced in that case by
\begin{align}
\text{Re}{\hat{\Sigma}}_{\tilde{d}_{11}}(m_{{\tilde{d}}_1}^2) &= 0.
\end{align}

The three mass-renormalization conditions determine essentially the 
counterterms for the diagonal mass parameters, $M_L^2$,
$M^2_{{\tilde{u}}_R}$ and $M^2_{{\tilde{d}}_R}$.
The non-diagonal counter\-terms $\delta \! A_u$ and $\delta \! A_d$ can be fixed
by imposing 
\begin{align}\label{A_Counterterm}
\text{Re}{\hat{\Sigma}}_{{\tilde{u}}_{12}}(m_{{\tilde{u}}_1}^2
) + \text{Re}{\hat{\Sigma}}_{{\tilde{u}}_{12}}(m_{{\tilde{u}}_2}^2)
&= 0  \\ 
\label{A_Countertermb}
\text{Re}{\hat{\Sigma}}_{{\tilde{d}}_{12}}(m_{{\tilde{d}}_1}^2
) + \text{Re}{\hat{\Sigma}}_{{\tilde{d}}_{12}}(m_{{\tilde{d}}_2}^2)
&= 0.
\end{align}

The diagonal $Z$-factors of the field-renormalization matrix \eqref{vollstTransfmatrix} 
can be determined by
the condition that the residues of the sfermion
propagators are unity,
\begin{align}\label{diagZ}
\text{Re}\frac{\partial {\hat{\Sigma}}_{{\tilde{f}}_{ii}}(k^2)}{\partial k^2} \biggl|_{k^2 =
  m_{\tilde{f}_{i}}^2} &= 0 \quad\ \text{for} \quad\ i=1,2 \quad\
  \text{and} \quad\ f=u,d\,.
\end{align}
There are two more
non-diagonal $Z$-factors of \eqref{vollstTransfmatrix} 
for each, $u$- and $d$-type, sfermion pair
at our disposal. They can be exploited to 
have zero mixing on each mass-shell. Imposing 
\begin{align}\label{nebendiagZ}
\text{Re}{\hat{\Sigma}}_{\tilde{f}_{12}}(m_{{\tilde{f}}_2}^2) &= 0
\quad\ \text{for} \quad\ f=u,d\, ,
\end{align}
yields, together with \eqref{A_Counterterm} and \eqref{A_Countertermb},
diagonal self-energies for each on-shell momentum $k^2$. 
Yet, one
$Z$-factor for each pair of sfermions remains 
undetermined. 
With the convenient choice
\begin{align}\label{nocheinZ}
\delta \! \breve{\mathcal{Z}}_{\tilde{f}_{12}} = \delta \!
\breve{\mathcal{Z}}_{\tilde{f}_{21}} \quad\ \text{for} \quad\ f=u,d
\end{align}
one obtains by
solving the equations \eqref{A_Counterterm}--\eqref{nocheinZ},
\begin{align}
{\delta \! \breve{\mathcal{Z}}_{\tilde{f}}}_{ii} &= - \text{Re} \frac{\partial
  {{\Sigma}}_{\tilde{f}_{ii}}(k^2)}{\partial k^2} \biggl|_{k^2 =
  m_{\tilde{f}_{i}}^2}\quad\ \text{for} \quad\ i=1,2 \quad\
  \text{and} \quad\ f=u,d\,, \\\label{nondiagZ}
\delta \! \breve{\mathcal{Z}}_{\tilde{f}_{12}} &= \delta \!
\breve{\mathcal{Z}}_{\tilde{f}_{21}} = -
  \frac{\text{Re}\Sigma_{\tilde{f}_{12}}(m_{{\tilde{f}}_{1}}^2) -
 \text{Re}\Sigma_{\tilde{f}_{12}}(m_{{\tilde{f}}_{2}}^2)
}{m_{{\tilde{f}}_{1}}^2 - m_{{\tilde{f}}_{2}}^2}
 \quad\ \text{for} \quad\ f=u,d\,.
\end{align}

For sleptons, the renormalization procedure can be applied analogously. 
Since we
have not introduced right-handed neutrinos, 
only the counter\-terms for the soft-breaking para\-meters
$M_{L_{\tilde{l}}}^2$, 
$M^2_{{\tilde{e}}_R}$ and $A_e$ have to be determined. Choosing the
conditions in analogy to \eqref{onshellBed1}, \eqref{onshellBed2} and
\eqref{A_Countertermb} we get: 
\begin{align}\label{onshellBed}
&\text{Re}{\hat{\Sigma}}_{\tilde{\nu}}(m_{{\tilde{\nu}}}^2) = 0\,,
\quad\ \text{Re}{\hat{\Sigma}}_{\tilde{e}_2}(m_{{\tilde{e}}_2}^2) = 0\,,
\\[2mm]\label{Ae_bedingung}
&\text{Re}{\hat{\Sigma}}_{{\tilde{e}}_{12}}(m_{{\tilde{e}}_1}^2
) + \text{Re}{\hat{\Sigma}}_{{\tilde{e}}_{12}}(m_{{\tilde{e}}_2}^2)
= 0\,.
\end{align} 

With the field and parameter renormalization constants determined
in the way described above, the 
renormalization of the sfermion sector is completed. The counter\-terms
are being implemented into the MSSM version of 
{\ttfamily  FeynArts} \nolinebreak \cite{HaScha} for completion at the
one-loop level.

\subsection{Determination of the renormalization constants}

The diagonal entries of the matrix~(\ref{vollstrenSelbstenergie})
of the renormalized self energies, for on-shell values of $k^2$,
are given by  $(i=1,2)$
\begin{align}
\label{diag}
\hat{\Sigma}_{\tilde{f}\,ii}(m_{{\tilde{f}}_{i}}^2)  &=
\Sigma_{\tilde{f}\,ii}(m_{{\tilde{f}}_{i}}^2) -
 \left( \mathcal{U}_{\tilde{f}}\,
 \delta \! \mathcal{M}_{\tilde{f}}\, \mathcal{U}^{+}_{\tilde{f}} 
 \right)_{ii} \nonumber \\[2mm]  
  &=
 \Sigma_{\tilde{f}\,ii}(m_{{\tilde{f}}_{i}}^2) - \;
 \delta  m_{\tilde{f}_i}^2 \, .
\end{align}
Solving the set of equations \eqref{onshellBed1} and \eqref{onshellBed2}
for the mass renormalization,
three out of the four squark-mass counterterms are determined
as follows,
\begin{align}
\delta  m_{\tilde{u}_1}^2 &=
\text{Re} \,\Sigma_{\tilde{u}_{11}}(m_{{\tilde{u}}_{1}}^2) \\[1mm]
\delta  m_{\tilde{u}_2}^2 &=
\text{Re} \,\Sigma_{\tilde{u}_{22}}(m_{{\tilde{u}}_{2}}^2) \\[1mm]
\delta  m_{\tilde{d}_2}^2 &=
\text{Re} \,\Sigma_{\tilde{d}_{22}}(m_{{\tilde{d}}_{2}}^2).
\end{align}
The fourth mass counter\-term is no
longer independent and can be 
expressed by the other counter\-terms of the soft-breaking para\-meters in the
following way,
\begin{align}\label{Countermassed1}
\begin{split}
\delta  m_{\tilde{d}_1}^2 &= U_{\tilde{d}_{11}}^2 \delta \!
M_{L_{\tilde{q}}}^2  + 2
U_{\tilde{d}_{11}}U_{\tilde{d}_{12}} \delta \! A_d +
U_{\tilde{d}_{12}}^2 \delta \! M_{{\tilde{d}}_R}^2 \\& \quad\ + U_{\tilde{d}_{11}}^2
\delta C_{\tilde{d}_{11}} +2 U_{\tilde{d}_{11}}U_{\tilde{d}_{12}} \delta C_{\tilde{d}_{12}} +
U_{\tilde{d}_{12}}^2 \delta C_{\tilde{d}_{22}} \, ,
\end{split}
\end{align}
with
\begin{align}
\begin{split}
\delta C_{\tilde{f}_{11}} &= 2 m_f \delta  m_f - Q_{{f}}
  M_Z^2 \cos(2 \beta)\, \delta \! \sin^2({\theta}_W)  \\[1.5mm]&\quad\ + \bigl(T_f^3 -
 Q_{{f}} \sin^2({\theta}_W)\bigr)\bigl( \cos(2 \beta)\, \delta \! M_Z^2 +
 M_Z^2\, \delta \! \cos(2 \beta)\bigr)\,, 
\end{split}\\[2mm]
\begin{split}\nonumber
 \delta C_{\tilde{f}_{12}} &= \delta C_{\tilde{f}_{21}} = \delta
  m_{f} ( A_{f} - \mu \kappa) - m_{f} \kappa \, \delta \! \mu - m_{f}\mu\,
 \delta  \kappa \\& \qquad\ \qquad\ \text{with} \quad\ \kappa =\left\{
 \begin{array}{r@{\quad} l} 
    \cot \beta & \text{for up-type squarks,} \; f = u\,, \\
\tan \beta & \text{for down-type squarks,}\; f = d\,, \end{array} \right.
\end{split} \\[2mm]
\begin{split}\nonumber
\delta C_{\tilde{f}_{22}} &= 2 m_f \delta  m_f +
 Q_{{f}}\bigl(M_Z^2 \cos(2 \beta)\, \delta \! \sin^2({\theta}_W) \\[1.5mm]&\quad\
 + \sin^2({\theta}_W) \cos(2 \beta)\, \delta \! M_Z^2 + M_Z^2
 \sin^2({\theta}_W)\, \delta \! \cos(2 \beta)\bigr)\,.
\end{split}
\end{align}
The counterterms $\delta \! M_{L_{\tilde{q}}}^2$, $\delta \!
M_{\tilde{u}_R,\tilde{d}_R}^2$, $\delta \! A_{u,d}$ 
for the basic parameters of the mass matrix
$\mathcal{M}_{\tilde{f}}$ in~(\ref{Sfermionmassenmatrix})
can be obtained from (\ref{diag})
and the non-diagonal entry 
$\delta Y_{\tilde{f}_{12}} = (\mathcal{U}_{\tilde{f}}
\delta \! \mathcal{M}_{\tilde{f}} \mathcal{U}^{+}_{\tilde{f}})_{12}$ 
as follows,
\begin{align}\label{deltaML}
\delta \! M_{L_{\tilde{q}}}^2 &= U_{\tilde{u}_{11}}^2
\delta  m_{\tilde{u}_1}^2 + U_{\tilde{u}_{12}}^2
\delta  m_{\tilde{u}_2}^2 - 2
U_{\tilde{u}_{12}}U_{\tilde{u}_{22}} \delta Y_{\tilde{u}_{12}}
 - \delta C_{\tilde{u}_{11}}\,,
 \\[4mm] 
 \delta \! M_{{\tilde{u}}_R}^2 &= U_{\tilde{u}_{12}}^2
 \delta  m_{\tilde{u}_1}^2 + U_{\tilde{u}_{11}}^2
 \delta  m_{\tilde{u}_2}^2 + 2
 U_{\tilde{u}_{12}}U_{\tilde{u}_{22}} \delta  Y_{\tilde{u}_{12}} - \delta C_{\tilde{u}_{22}}
\\[4mm]  \label{deltaMdR}
\delta \! M_{{\tilde{d}}_R}^2 &= \frac{U_{\tilde{d}_{11}}^2 - U_{\tilde{d}_{12}}^2}{U_{\tilde{d}_{11}}^2}
\delta  m_{\tilde{d}_2}^2 + 2
\frac{U_{\tilde{d}_{12}} U_{\tilde{d}_{22}}}{U_{\tilde{d}_{11}}^2}
\delta Y_{\tilde{d}_{12}} +
\frac{U_{\tilde{d}_{12}}^2 U_{\tilde{u}_{11}}^2}{U_{\tilde{d}_{11}}^2}
\delta  m_{\tilde{u}_1}^2  \\[1.5mm]&
\nonumber \quad\
+ \frac{U_{\tilde{d}_{12}}^2  U_{\tilde{u}_{12}}^2}{U_{\tilde{d}_{11}}^2}
\delta  m_{\tilde{u}_2}^2 - 2 \frac{U_{\tilde{d}_{12}}^2 U_{\tilde{u}_{12}} U_{\tilde{u}_{22}}}{U_{\tilde{d}_{11}}^2}
\delta Y_{\tilde{u}_{12}}
 - \delta C_{\tilde{d}_{22}}  +
\frac{U_{\tilde{d}_{12}}^2}{U_{\tilde{d}_{11}}^2} (\delta 
C_{\tilde{d}_{11}} - \delta C_{\tilde{u}_{11}})\,,
 \\[4mm]\label{deltaAu}
 \delta \! A_u &= \frac{1}{m_u}\Bigl[U_{\tilde{u}_{11}}
 U_{\tilde{u}_{12}}\bigl(\delta  m_{\tilde{u}_1}^2 
 - \delta  m_{\tilde{u}_2}^2\bigr)
 + (U_{\tilde{u}_{11}} U_{\tilde{u}_{22}} 
 +U_{\tilde{u}_{12}}
 U_{\tilde{u}_{21}}) \delta Y_{\tilde{u}_{12}}
  - \delta C_{\tilde{u}_{12}} \Bigr]
 \\[4mm]\label{deltaAd} 
\delta \! A_d &= \frac{1}{m_d}\Bigl[
-\frac{U_{\tilde{d}_{12}}}{U_{\tilde{d}_{11}}} 
\delta  m_{\tilde{d}_2}^2 +
\frac{U_{\tilde{d}_{22}}}{U_{\tilde{d}_{11}}} \delta Y_{\tilde{d}_{12}} 
+ \frac{U_{\tilde{d}_{12}} U_{\tilde{u}_{11}}^2}{U_{\tilde{d}_{11}}}
\delta  m_{\tilde{u}_1}^2
\\[1.5mm]& \nonumber \quad\ \quad\ + \frac{U_{\tilde{d}_{12}}
  U_{\tilde{u}_{12}}^2}{U_{\tilde{d}_{11}}} 
\delta  m_{\tilde{u}_2}^2 - 2
\frac{U_{\tilde{d}_{12}} U_{\tilde{u}_{12}} U_{\tilde{u}_{22}}}{ 
  U_{\tilde{d}_{11}}} \delta Y_{\tilde{u}_{12}}
- \delta C_{\tilde{d}_{12}} +
\frac{U_{\tilde{d}_{12}}}{U_{\tilde{d}_{11}}} (\delta 
C_{\tilde{d}_{11}} - \delta C_{\tilde{u}_{11}})\Bigr]\, .
\end{align} 
$\delta Y_{\tilde{f}_{12}}$
is determined with the help of
(\ref{A_Counterterm}), (\ref{A_Countertermb}), (\ref{nondiagZ})
to be
\begin{align}
\label{y12}
\delta Y_{\tilde{f}_{12}} = \frac{1}{2}\bigl(\text{Re}{{\Sigma}}_{{\tilde{f}}_{12}}(m_{{\tilde{f}}_1}^2
) + \text{Re}{{\Sigma}}_{{\tilde{f}}_{12}}(m_{{\tilde{f}}_2}^2)\bigr) \quad\ \text{for} \quad f = u, d\,.
\end{align}
Inserting the expressions for $\delta \!M_{L_{\tilde{q}}}^2$,
$\delta \! M_{{\tilde{d}}_R}^2$ 
and $\delta \! A_d$ 
into \eqref{Countermassed1}, the mass
counterterm $\delta  m_{\tilde{d}_1}^2$ can be written as
\begin{align}
\begin{split}\label{dmd1Counterterm}
\delta  m_{\tilde{d}_1}^2 &= -
\frac{U_{\tilde{d}_{12}}^2}{U_{\tilde{d}_{11}}^2} \delta 
m_{\tilde{d}_2}^2 + 2 \frac{ U_{\tilde{d}_{12}}
  U_{\tilde{d}_{22}}}{U_{\tilde{d}_{11}}^2}\delta Y_{\tilde{d}_{12}} 
+\frac{U_{\tilde{u}_{11}}^2}{U_{\tilde{d}_{11}}^2} \delta 
m_{\tilde{u}_1}^2   +\frac{U_{\tilde{u}_{12}}^2}{U_{\tilde{d}_{11}}^2} 
\delta  m_{\tilde{u}_2}^2  
\\& \quad\ 
- 2 \frac{U_{\tilde{u}_{12}}
  U_{\tilde{u}_{22}}}{U_{\tilde{d}_{11}}^2}\delta Y_{\tilde{u}_{12}} + 
\frac{1}{U_{\tilde{d}_{11}}^2}( \delta C_{\tilde{d}_{11}}   - \delta 
C_{\tilde{u}_{11}})\,.
\end{split}
\end{align}
This relation contains, besides those
counterterms determined within the sfermion sector,
also renormalization constants that have to be taken from other sectors:
the fermion-mass counterterm $\delta  m_f$, 
the gauge-boson mass counterterms $\delta \! M_{W,Z}^2$,
and $\delta \! \tan \beta$.   
The renormalization of the electroweak mixing angle,  $\delta\sin^2 \theta_W$, 
follows from the relation 
$\sin^2\theta_{W} = 1 - \frac{M_W^2}{M_Z^2}$ 
(actually, in the combination of (\ref{dmd1Counterterm}),   
$\delta \! M_Z^2$ drops out).
$\delta \mu$ can be obtained from renormalization in the
chargino sector and is given explicitly in \cite{fh};
$\delta m_{\tilde{d}_1}^2$ is, however,
independent of $\delta C_{\tilde{f}_{12}}$ and hence also independent
of $\delta \mu$.

\medskip \noindent
The counter\-terms $\delta  m_f$, $\delta \! \tan \beta$ and $\delta
\!M_{W,Z}^2$ are determined from the the following conditions. 
\begin{itemize}
\item[(i)]
On-shell renormalization of the  fermion mass yields
\nolinebreak \cite{De} 
\begin{align}
\delta  m_f = \frac{1}{2} m_f \bigl (\text{Re}{\Sigma}_{{f}_L} (m_f^2) +
\text{Re} {\Sigma}_{{f}_R} (m_f^2) + 2 \text{Re}{\Sigma}_{{f}_S} (m_f^2) \bigr)
\end{align} 
in terms of the fermion self energy
\begin{align}
\Sigma_f(k) = {\Sigma}_{{f}_L}(k^2)\! \not{k} {\rm \boldmath P_L}
              + {\Sigma}_{{f}_R}(k^2)\! \not{k} {\rm \boldmath P_R}
              + m_f {\Sigma}_{{f}_S}(k^2) \, .
\end{align}
\item[(ii)]
On-shell renormalization of the gauge-boson masses determines the
mass counter\-terms 
\begin{align}
\delta \! M_V^2 = \text{Re}{\Sigma}_{\text{V}}(M_V^2) 
       \quad \quad {\rm for} \quad V=W,Z \, ,
\end{align}
in terms of the vector-boson self energies ${\Sigma}_{\text{V}}(k^2)$.
\item[(iii)]
Vanishing $A^0$-$Z$-mixing for an on-shell $A^0$-boson 
determines the counter\-term of $\tan \beta$ according to \nolinebreak \cite{Da}  
\begin{align}
\label{tanbetaren}
\delta \! \tan \beta =  
\frac {1}{2 M_Z \cos^2 \beta} \text{Im}{\Sigma}_{A^0 Z}(M^2_{A}).
\end{align}
Another option is to renormalize $\tan \beta$ in the
$\overline{DR}$-scheme \cite{franketal}
where only the UV-singular part of (\ref{tanbetaren}) is 
taken into account, which has the advantage of avoiding
large finite contributions and providing a
gauge invariant and process independent
counterterm~\cite{FreiStoe}. 
Comparing both renormalization schemes, 
the numerical results for the sfermion masses differ by
at most $\mathcal O(10\;\text{MeV})$. A potentially 
large finite part of (\ref{tanbetaren}) for large values of
$\tan \beta$ is suppressed  by
a factor $\tan \beta/(1 + \tan^2 \beta)^2$ in the sfermion-mass 
counterterms, which keeps the result stable.
\end{itemize}

After this specification of all the renormalization constants the 
squark sector is completed at the one-loop level.
Another way of renormalization, performed in \cite{GuSoHo},
consists of introducing counterterms directly for the physical parameters, 
i.e.\ masses and mixing angles, 
instead of the soft-breaking parameters.
In that case also the transformation matrix 
${\mathcal{U}}_{\tilde{f}}$ in (\ref{transfmatrix})
has to be renormalized by the mixing angle counterterm,
i.e.\ through 
$\theta_{\tilde{f}}\rightarrow \theta_{\tilde{f}}
                             + \delta\theta_{\tilde{f}}$,
whereas in our case ${\mathcal{U}}_{\tilde{f}}$ is not affected.   
Previous studies of the sfermion mass spectrum \cite{Pierceetal} were
done in the $\overline{DR}$-scheme with running parameters whereby
the MSSM parameter space is restricted by unification assumptions.

\medskip 
The treatment of one generation of sleptons is similar to the one of 
squarks. Two of the three slepton masses are fixed 
by on-shell conditions, and the third one is dependent on
the other counter\-terms,
\begin{align}\label{dme1}
\begin{split}
\delta  m_{\tilde{e}_1}^2 &= U_{\tilde{e}_{11}}^2 \delta \!
M_{L_{\tilde{l}}}^2 + 2 
U_{\tilde{e}_{11}}U_{\tilde{e}_{12}} \delta \! A_e +
U_{\tilde{e}_{12}}^2 \delta \! M_{{\tilde{e}}_R}^2 \\& \quad\ + U_{\tilde{e}_{11}}^2
\delta C_{\tilde{e}_{11}} +2 U_{\tilde{e}_{11}}U_{\tilde{e}_{12}} \delta C_{\tilde{e}_{12}} +
U_{\tilde{e}_{12}}^2 \delta C_{\tilde{e}_{22}}.
\end{split}
\end{align}
The quantities  $\delta \! M_{L_{\tilde{l}}}^2$, $\delta \! M_{{\tilde{e}}_R}^2$
and $\delta \! A_e$ follow from  equations \eqref{onshellBed},
\eqref{Ae_bedingung} 
and are explicitly given by
\begin{align}
\delta \! M_{L_{\tilde{l}}}^2  &= \delta  m_{\tilde{\nu}}^2 - \delta C_{\tilde{\nu}}\,,
\\[4mm]  \label{deltaMR}
\delta \! M_{{\tilde{e}}_R}^2 &= \frac{U_{\tilde{e}_{11}}^2 -
  U_{\tilde{e}_{12}}^2}{U_{\tilde{e}_{11}}^2} 
\delta  m_{\tilde{e}_2}^2 + 2
\frac{U_{\tilde{e}_{12}} U_{\tilde{e}_{22}}}{U_{\tilde{e}_{11}}^2}
\delta Y_{\tilde{e}_{12}} +
\frac{U_{\tilde{e}_{12}}^2}{U_{\tilde{e}_{11}}^2}
\delta  m_{\tilde{\nu}}^2  
 - \delta C_{\tilde{e}_{22}}  +
\frac{U_{\tilde{e}_{12}}^2}{U_{\tilde{e}_{11}}^2} (\delta 
C_{\tilde{e}_{11}} - \delta C_{\tilde{\nu}})\,,
 \\[4mm]\label{deltaAe} 
\delta \! A_e &= \frac{1}{m_e}\Bigl[
-\frac{U_{\tilde{e}_{12}}}{U_{\tilde{e}_{11}}} 
\delta  m_{\tilde{e}_2}^2 +
\frac{U_{\tilde{e}_{22}}}{U_{\tilde{e}_{11}}} \delta Y_{\tilde{e}_{12}} 
+ \frac{U_{\tilde{e}_{12}}}{U_{\tilde{e}_{11}}}
\delta  m_{\tilde{\nu}}^2
- \delta C_{\tilde{e}_{12}} +
\frac{U_{\tilde{e}_{12}}}{U_{\tilde{e}_{11}}} (\delta
C_{\tilde{e}_{11}} - \delta C_{\tilde{\nu}})\Bigr]\,,
\end{align} 
with
\begin{align}
\delta Y_{\tilde{e}_{12}} &= \frac{1}{2}\bigl(\text{Re}{{\Sigma}}_{{\tilde{e}}_{12}}(m_{{\tilde{e}}_1}^2
) + \text{Re}{{\Sigma}}_{{\tilde{e}}_{12}}(m_{{\tilde{e}}_2}^2)\bigr) \,, 
\\
\delta C_{\tilde{\nu}} &=  \frac{1}{2} \bigl(\cos(2 \beta) \delta \! M_Z^2 +
 M_Z^2 \delta \! \cos(2 \beta)\bigr)\,, 
\\[2mm]
\delta C_{\tilde{e}_{11}} &= 2 m_e \delta  m_e +
  M_Z^2 \cos(2 \beta) \delta \! \sin^2({\theta}_W)  
\\[1.5mm]&\quad\ \nonumber
- (\frac{1}{2} - \sin^2({\theta}_W))\bigl( \cos(2 \beta) \delta \! M_Z^2 +
 M_Z^2 \delta \! \cos(2 \beta)\bigr)\,, 
\\[2mm]
 \delta C_{\tilde{e}_{12}} &= \delta C_{\tilde{e}_{21}} = \delta
  m_{e} ( A_{e} - \mu \tan \beta) - m_{e}\tan \beta \,\delta \! \mu - m_{e}\mu\,
 \delta \! \tan \beta \,,
 \\[2mm]
\delta C_{\tilde{e}_{22}} &= 2 m_e \delta  m_e -
 M_Z^2 \cos(2 \beta) \delta \! \sin^2({\theta}_W) 
 - \sin^2({\theta}_W) \cos(2 \beta) \delta \! M_Z^2 
\\[2mm]& \quad \nonumber
- M_Z^2
 \sin^2({\theta}_W) \delta \! \cos(2 \beta)\,.
\end{align}
These expressions complete the renormalization also in the slepton sector.

\subsection{Mass corrections}

The sfermion masses fixed via the on-shell conditions
\eqref{onshellBed1}, \eqref{onshellBed2} for squarks and 
\eqref{onshellBed} for sleptons
do not receive any corrections at 
one-loop order. The remaining mass, in each squark or slepton generation,
is different at tree level and one-loop order. 
The counter\-term~\eqref{Countermassed1} 
or \eqref{dme1}, respectively,
absorbs the divergence of the corresponding self 
energy, but it leaves a  finite contribution. The shifts $\Delta
m_{\tilde{d}_1}^2$ and $\Delta m_{\tilde{e}_1}^2$ for the pole masses are
given by
\begin{align}
\Delta m_{\tilde{d}_1}^2 = \delta m_{\tilde{d}_1}^2 -
\text{Re} {{\Sigma}}_{\tilde{d}_{11}}(m_{\tilde{d}_{1}}^2) \quad\
\text{and} \quad\ \Delta m_{\tilde{e}_1}^2 = \delta m_{\tilde{e}_1}^2 -
\text{Re} {{\Sigma}}_{\tilde{e}_{11}}(m_{\tilde{e}_{1}}^2),  
\end{align}  
yielding  one-loop masses
according to
\begin{align}
m_{{{\tilde{d}}_1\text{1-Loop}}}^2 =
{m_{\tilde{d}_{1_{\text{Born}}}}^2 + \Delta m_{\tilde{d}_1}^2} \quad\
\text{and} \quad\ 
m_{{{\tilde{e}}_1\text{1-Loop}}}^2 =
{m_{\tilde{e}_{1_{\text{Born}}}}^2 + \Delta m_{\tilde{e}_1}^2},
\end{align}
where $m_{\tilde{d}_{1_{\text{Born}}}}$ and
$m_{\tilde{e}_{1_{\text{Born}}}}$ are the masses
in the Born approximation. 
In the self energies ${{\Sigma}}_{\tilde{d}_{11}}$ and
${{\Sigma}}_{\tilde{e}_{11}}$,
the masses can be taken as the lowest order masses.

\section{Numerical results and discussion}
\label{numres}

\psfrag{Mdschl1}{\begin{rotate}{180}{$m_{{\tilde{d}}_1}$
    [GeV]}\end{rotate}} 
\psfrag{Msschl1}{\begin{rotate}{180}{$m_{{\tilde{s}}_1}$
    [GeV]}\end{rotate}}
\psfrag{mbschl1}{\begin{rotate}{180}{$m_{{\tilde{b}}_i}$
    [GeV]}\end{rotate}}
\psfrag{Mbschl1}{\begin{rotate}{180}{$m_{{\tilde{b}}_1}$
    [GeV]}\end{rotate}}
\psfrag{Meschl1}{\begin{rotate}{180}{$m_{{\tilde{e}}_1}$
    [GeV]}\end{rotate}}
\psfrag{Mmuschl1}{\begin{rotate}{180}{$m_{{\tilde{\mu}}_1}$
    [GeV]}\end{rotate}}
\psfrag{Mtauschl1}{\begin{rotate}{180}{$m_{{\tilde{\tau}}_1}$
    [GeV]}\end{rotate}}
\psfrag{Au}{\raisebox{-0.8ex}{$A_u$ [GeV]}}
\psfrag{Ad}{\raisebox{-0.8ex}{$A_d$ [GeV]}}
\psfrag{tanbeta}{\raisebox{-0.5ex}{{$\tan \beta$}}}
\psfrag{Massenkorrekturstark}{}
\psfrag{Massenkorrekturelschwach}{} 
\psfrag{EinschleifenMasse}{}
\psfrag{BornMasse}{}
\begin{figure}[!t]
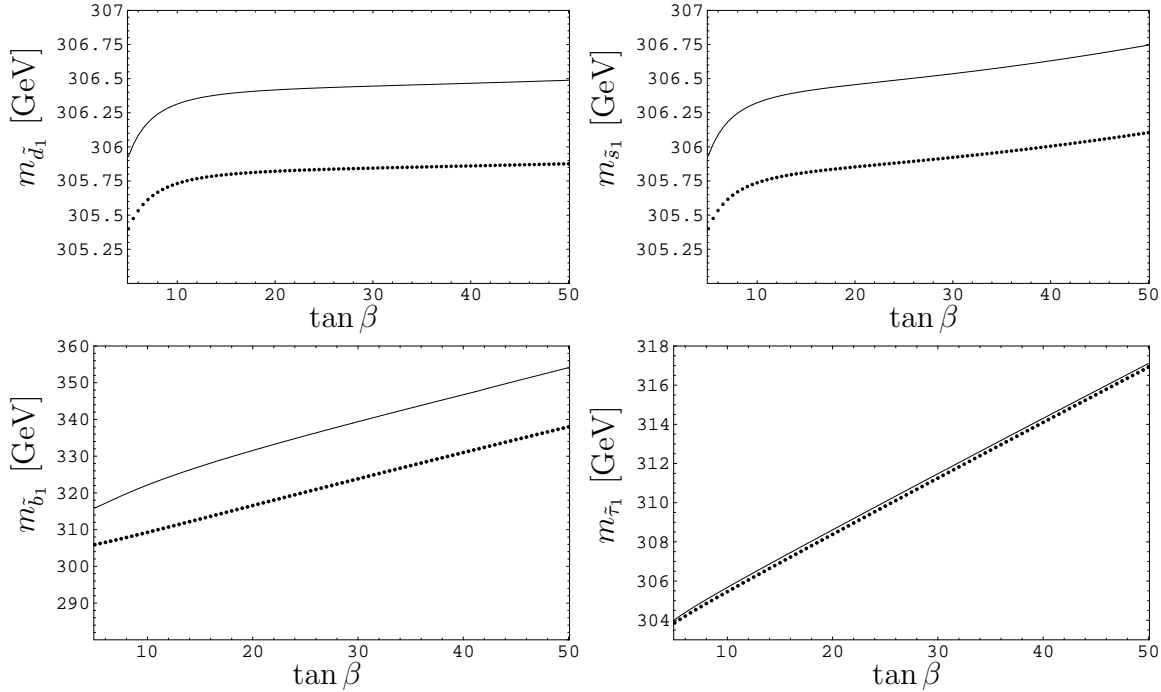

\begin{tabular}{cc}
\includegraphics[width=0.45\linewidth]{fig.1a} &
\includegraphics[width=0.45\linewidth]{fig.1b} \\
\includegraphics[width=0.45\linewidth]{fig.1c} &
\includegraphics[width=0.45\linewidth]{fig.1d}
\end{tabular}
\caption{\sf Born (dotted line) and one-loop masses (solid line) as a function of
  $\tan \beta$. In the 
  first row the masses of the $\tilde{d}_1$-squark and the $\tilde{s}_1$-squark
  are shown, in the second row the masses of the $\tilde{b}_1$-squark
  and the $\tilde{\tau}_1$-slepton. The para\-meters have been
  chosen as $M_{\tilde{f}_R} = M_L = A_f = 300$ GeV for all generations.
\label{tanbetaAbh} }
\end{figure}

The self energies  were calculated with the help of the programs {\ttfamily
  FeynArts}, {\ttfamily FormCalc} and {\ttfamily
  LoopTools} \nolinebreak \cite{HaScha, Ha}, with 
  the method of  
\ "Constrained Differential Renormalization" \nolinebreak \cite{AgHa}
  for regularization. This method is equivalent to the procedure of 
  dimensional reduction \nolinebreak \cite{Si}.

\medskip
In the following, we illustrate the effect of the one-loop contributions
for specific examples in Figures \ref{tanbetaAbh} to
\ref{AAbh}.  Unless stated otherwise,
the default values for the parameters
listed in appendix \ref{Parameter} are used.

\begin{figure}[!b]
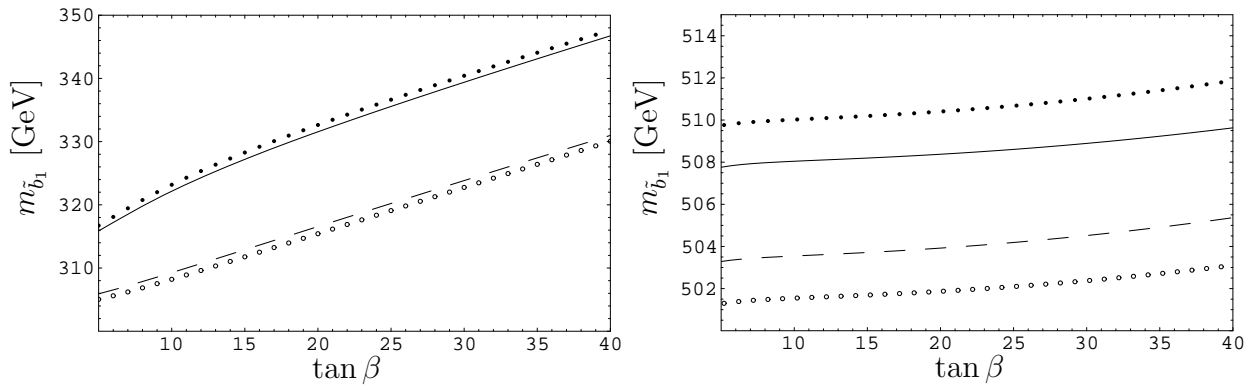

\begin{tabular}{ll}
\includegraphics[width=.485\linewidth]{fig.2a}&
\includegraphics[width=.485\linewidth]{fig.2b}
\end{tabular}\caption{\sf Total one-loop mass (solid line) of the
  $\tilde{b}_1$-particle 
  as a function of $\tan \beta$ in comparison to the Born mass (dashed
  line) and the
  one-loop masses with either strong (dotted line) and electroweak
  (line of circles)
  contributions. The soft breaking para\-meters are chosen as
  $M_{L_{\tilde{q}_3}} = 300$ GeV in the left and
  $M_{L_{\tilde{q}_3}} = 500$ GeV in the right figure,
   and  $M_{\tilde{f}_R}  = M_{L_{\tilde{f} \neq
    \tilde{q}_3}} = A_f = 300$ GeV in both.  
\label{Einschleifenmassen} }
\end{figure}

\medskip
The size of the mass shift for the three  squark generations
is displayed in Figure \ref{tanbetaAbh}, together with the
correction to the $\tilde{\tau}_1$ mass as an example for the sleptons.
Because of the presence of the fermion mass in the off-diagonal entry 
of \eqref{Sfermionmassenmatrix}, the dependence of the  
sfermion masses on $\tan\beta$ is strongest
for the third generation.
The mass shifts are nearly independent of $\tan \beta$
for all the particles. They are rather small
(0.6 GeV)
in the first two squark generations,
but they are much larger in the third generation.
The Born mass of the
$\tilde{b}_1$-squark is  enhanced significantly 
by up to 16 GeV (5\%) at the one-loop level.
The mass shift for sleptons is of electroweak origin only
and is hence rather small, for the
$\tilde{\tau}_1$-slepton only 0.2 GeV (or 0.1\%).

\medskip
The various contributions to the one-loop mass shift versus $\tan \beta$
are shown in Figure~\ref{Einschleifenmassen}, 
for the case of the $\tilde{b}_1$ squark: 
the Born mass and the one-loop mass,
together with the individual parts from the  
strong and the electroweak interactions.
The biggest shift originates from the strong 
interaction, i.e.\ by virtual squarks, gluinos, quarks and gluons,  and
amounts to approximately 17 GeV (5\%)
and 6.5 GeV (1.5\%) for
$M_{L_{\tilde{q}_3}} = 300$ GeV and $M_{L_{\tilde{q}_3}} = 500$ GeV,
respectively. 

\medskip
The electroweak contribution 
can become also more sizeable, as the example of the right part in 
Figure~\ref{Einschleifenmassen} shows, with
$M_{L_{\tilde{q}_3}} = 500$ GeV,
where a shift of 2.3 GeV is observed.
The electroweak
contributions result from virtual sleptons, squarks, charginos, neutralinos
and quarks, Higgs-, W- and Z-bosons and photons. 
Since the strong and
the electroweak contributions have opposite signs,
the total correction adds up to 5\% (16 GeV) and 1\% (4.2 GeV) of the
Born mass for
$M_{L_{\tilde{q}_3}} = 300$ GeV and $M_{L_{\tilde{q}_3}} = 500$ GeV,
respectively.

\begin{figure}[!t]
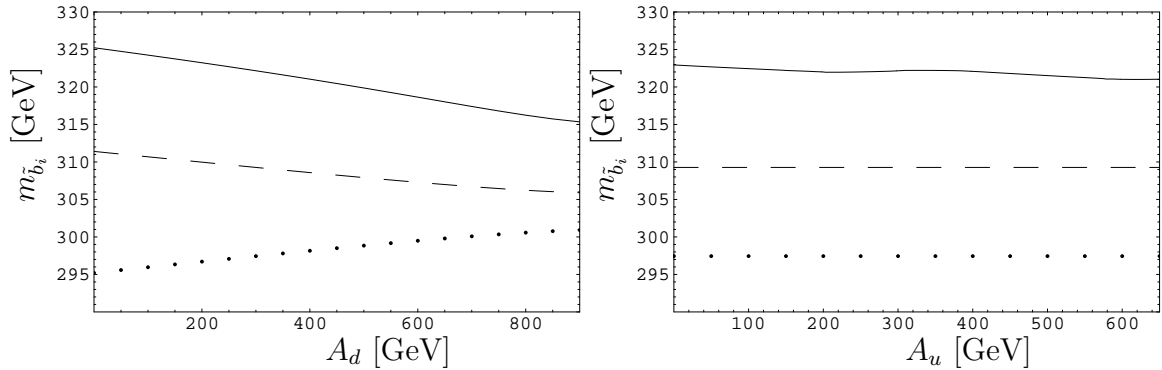

\begin{tabular}{cc}
\includegraphics[width=0.45\linewidth]{fig.3a} &
\includegraphics[width=0.45\linewidth]{fig.3b} 
\end{tabular}
\caption{\sf Born (dashed line) and one-loop masses (solid line) of the
  $\tilde{b}_1$-squark and the mass 
  of the $\tilde{b}_2$-squark (dotted line) as a
  function of the $A$-para\-meter $A_d$ or $A_u$. 
  The para\-meters have been 
  chosen as $M_{\tilde{f}_R} = M_L = 300$ GeV for all generations. It is
  assumed that $A_d = A_s = A_b$ and $A_u = A_c = A_t$. When $A_d$ is
  varied then $A_u = 300$ GeV and vice versa.
\label{AAbh}}
\end{figure}

Finally, the dependence on the $A$-para\-meters is considered. 
Exemplary, the $A$-para\-meter dependence is shown for the
$\tilde{b}_i$-squarks in Figure \ref{AAbh}.
Varying the para\-meter $A_d$ changes the mixing of the
bottom squarks, an effect which is suppressed by $m_f$ in the 
light generations.
The corrections to the Born mass show a weak dependence on the para\-meter
$A_d$. They decrease from 14 GeV to 9 GeV in the $A_d$ range of
Figure~\ref{AAbh}. 

The Born
masses of the bottom squarks do not depend on the para\-meter $A_u$, but one
can see a slight decrease of the corrections to the mass of $\tilde{b}_1$ 
when $A_u$ is increased. $A_u$
changes the mixing and the mass splitting of the up-type squarks, which
influences slightly the size of the mass shift.

\section{Conclusion}
We have presented a complete on-shell renormalization of the 
scalar-fermion sector of the MSSM
based on the entire set of one-loop diagrams,
treating all masses as pole masses 
and with renormalization constants that allow to formulate the sfermion self-energies
as matrices which become diagonal for external momenta on-shell.
The renormalization conditions are specified to fix the counter\-terms
of the basic soft-breaking
para\-meters, respecting $SU(2)$-invariance.
As an application, we have calculated the sfermion-mass spectrum at the one-loop
level.
Three of the four
squark tree-level  masses and two of the three slepton tree-level
masses  can be made equal to the corresponding
one-loop pole masses. The residual squark and slepton mass, instead,
receives a mass shift at one-loop level.
These mass shifts are rather small for sleptons,
but they can be sizeable, of the order of 5\%, for squarks. 
Thus, especially for the third  generation, this
mass shift has to be taken into account in precision calculations.

\newpage

\section*{Acknowledgement}
We want to thank
D. St\"ockinger for useful discussions.
This work was supported in part 
by the European Community's Human Potential Programme under contract
HPRN-CT-2000-00149 "Physics at Colliders".

\vspace*{2cm}

\section*{Appendix}

\begin{appendix}
\section{Parameters}\label{Parameter}

If not mentioned explicitly in the text, the following default set of
parameters is used:
\begin{itemize}
\item parameters of the Higgs sector:\\[1.0ex]
\renewcommand{\arraystretch}{1.5}
\begin{tabular}{lclcl}
 $M_A = 150$ GeV &\hspace{1.5cm} &
  $\tan \beta = 10$ &\hspace{1.5cm} & $\mu = 100$ GeV \\ (Mass of the $A^0$-boson) & & & & 
\end{tabular}
\item soft-breaking parameters:\\[1.0ex] \nopagebreak
\renewcommand{\arraystretch}{1.5}
\begin{tabular}{lcl}
 for the gauginos:& \hspace{2cm} & for the sfermions:
\\[2mm]  $M_1 = \displaystyle{\frac{5}{3} \frac{\sin^2
  \theta_W}{\cos^2 \theta_W}} M_2$ & &  $M_L =
M_{L_{\{\tilde{q}_i,\,\tilde{l}_i\}}} = 300$ GeV with $i = 1,\, 2,\,3$\\
 $M_2 = 200$ GeV & & $M_{\tilde{f}_R} = 300$ GeV with $f =
 u,\,c,\,t,\,d,\,s,\,b,\,e,\,\mu,\,\tau$\\
 $M_3 =\displaystyle{\frac{\alpha_s}{\alpha}}{\sin^2 \theta_W}M_2$ & &
 $A_{\{u,\,c,\,t\}} = A_{\{d,\,s,\,b\}} = A_{\{e,\,\mu,\,\tau\}} = 300$ GeV 
\end{tabular}
\end{itemize}
\end{appendix}

\end{document}

%% file: titlepage.tex
\thispagestyle{empty}
\setcounter{page}{0}
\def\thefootnote{\fnsymbol{footnote}}

{\textwidth 16cm

\begin{flushright}
MPI-PhT/2003-20 \\
hep-ph/0305328 \\
\end{flushright}

\vspace{2cm}

\begin{center}

{\large\sc {\bf The  sfermion mass spectrum of the MSSM\\[0.3cm]
     at the one-loop level}}

\vspace{2cm}

{\sc W. Hollik} and {\sc H. Rzehak}

\vspace*{0.4cm}
    
      Max-Planck-Institut f\"ur Physik, F\"ohringer Ring 6 \\
      D-80805 M\"unchen, Germany

\end{center}

\vspace*{2cm}

\begin{abstract}
\noindent
The sfermion-mass spectrum of the minimal supersymmetric standard model is 
investigated at the one-loop level. An on-shell scheme has been specified
for renormalization of the basic breaking parameters of the sfermionic 
sector. Owing to $SU(2)$-invariance, the soft-breaking 
mass parameters of the left-chiral scalar fermions
of each isospin doublet are identical. 
Thus, one of the 
sfermion-masses  of each doublet can be expressed in terms of the
other masses and receives a mass shift at the one-loop level 
with respect to the lowest-order value, which can be of $O(10$ GeV). 
Both strong and  electroweak contributions have been calculated
for scalar quarks and leptons.
\end{abstract}

}
\def\thefootnote{\arabic{footnote}}
\setcounter{footnote}{0}

\newpage